\documentclass[%
 reprint,
 amsmath,amssymb,
 aps, prl
]{revtex4-2}

\usepackage[T1]{fontenc}
\usepackage[utf8]{inputenc}
\usepackage{babel}
\usepackage{graphicx}% Include figure files
\usepackage{dcolumn}% Align table columns on decimal point
\usepackage{bm}% bold math
\usepackage{enumerate}
\usepackage{amssymb}
\usepackage{amsbsy}
\usepackage{bm}
\usepackage{mathtools}
\usepackage{physics}
\usepackage{siunitx}
\usepackage{bbm}
\usepackage{abraces}
\usepackage{mathrsfs}
\usepackage{xcolor}
\usepackage{mathrsfs}
\usepackage{bm}
\usepackage{hyperref}
\usepackage[capitalise]{cleveref}

\newcommand{\im}{\mathrm{i}}
\newcommand{\up}{\uparrow}
\newcommand{\down}{\downarrow}
\newcommand{\expec}[1]{\langle #1 \rangle}

\newcommand{\vk}{\bm{k}}
\newcommand{\vq}{\bm{q}}

\ExplSyntaxOn
\newcommand{\processmomenta}[1]{%
  \tl_set:Nn \l_tmpa_tl {#1}%
  \tl_replace_all:Nnn \l_tmpa_tl {k}{\vk}%
  \tl_replace_all:Nnn \l_tmpa_tl {q}{\vq}%
  \tl_use:N \l_tmpa_tl%
}
\ExplSyntaxOff

\NewDocumentCommand\ladder{O{}mm}{%
  \hat{c}_{%
    \processmomenta{#2},%
 #3%
 }^{%
    \IfNoValueTF{{#1}}{{\phantom{\dagger}}}{#1}%
 }%
}

\newcommand{\opSC}[1]{\ladder{-#1}{\down} \ladder{#1}{\up}}
\newcommand{\opSCd}[1]{\ladder[\dagger]{#1}{\up} \ladder[\dagger]{-#1}{\down}}

\newcommand{\Ef}{E_\mathrm{F}}
\newcommand{\rf}{\rho_\mathrm{F}}
\newcommand{\ep}{\varepsilon_\mathrm{peak}}
\newcommand{\vp}{\varepsilon}

\newcommand{\debye}{\omega_\mathrm{D}}
\newcommand{\kb}{k_\mathrm{B}}

\newcommand{\maxGap}{\Delta_\mathrm{max}}
\newcommand{\trueGap}{\Delta_\mathrm{true}}

\newcommand{\greenMatSymbol}{\mathcal{G}}

\newcommand{\spectralMatSymbol}{\mathcal{A}}
\newcommand{\spectralMat}[1]{\spectralMatSymbol_\mathrm{#1}  (\omega)}

\newcommand{\md}{\mathrm{d}}
\newcommand{\hamiltonian}{\mathcal{H}}

\begin{document}

%\preprint{APS/123-QED}

\title{Enhanced Superconductivity in Proximity to Peaks in Densities of States}% Force line breaks with \\

\author{Joshua Alth\"user}
\email{joshua.althueser@tu-dortmund.de}
\author{Ilya M. Eremin}
\email{ilya.eremin@ruhr-uni-bochum.de}
\author{G\"otz S. Uhrig}%
\email{goetz.uhrig@tu-dortmund.de}
\affiliation{TU Dortmund University, Otto-Hahn Stra\ss{}e 4, 44227 Dortmund, Germany}
\affiliation{Theoretische Physik III, Ruhr-Universit\"at Bochum, 44801 Bochum, Germany}
\date{\today}% It is always \today, today,

\begin{abstract}
For the BCS theory of superconductivity, the electron-phonon interaction is transformed to an attractive 
electron-electron interaction in the vicinity of the
Fermi energy only. At the same time, its formal derivation using a unitary transformation 
reveals that the electrons attract one another whenever their energies do
not differ more than the phonon energy $\debye$, independent of closeness to the Fermi energy. Consequently, the order parameter becomes finite even away from the Fermi level. Yet, for small interactions, its magnitude is usually small and can be safely ignored, justifying the BCS approximation.
Intriguingly, we find that an accumulation of density-of-states at an energy $\ep$ in proximity to the Fermi energy induces a significant order parameter magnitude around $\ep$, which exceeds the one at $\Ef$ for moderate coupling strengths. 
This strong enhancement is heralded by the softening of an additional collective mode, which resembles a second phase transition. We predict measurable signatures in the thermodynamic and spectroscopic response of this unexpected phenomenon, guiding future experimental searches for it.
\end{abstract}

\maketitle

%\tableofcontents

%%%%%
% Intro
%%%%%
The weak-coupling regime of superconductors is well-understood in the framework of the BCS theory\cite{bardeen1957}.
The essential result is that the superconducting instability is determined by the product of the attractive coupling times the electronic density of states (DOS) at the Fermi energy, and the transition temperature becomes higher by increasing at least one of them. The relevant physics is given by what happens in an energetically narrow layer around the Fermi energy, and its width is given by the characteristic phononic Debye frequency $\debye$.
Since then, strong-coupling superconductors have been a central focus of research for decades and continue to incite great
interest \cite{eliashberg1960,scalapino1966,leavens1977,combescot1995,joas2002,chubukov2020,yuzbashyan2022,tajima2024,lorenzana2024a,althuser2025},
striving to realize high-T$_c$ superconductivity and the limits beyond the pure weak-coupling scenario.

In conventional superconductors, the attractive electron-electron interaction is derived from the electron-phonon coupling 
by a unitary transformation \cite{frohlich1952,lenz1996,mielke1997,mielke1997a,kehrein2006} if one does not resort to an explicit description of
the phonon degrees of freedom as well \cite{eliashberg1960}. 
The {standard weak-coupling} form of the attraction between an electron at
energy $\varepsilon$ and one at energy $\varepsilon'$ reads
\begin{align}
g_\mathrm{BCS}(\varepsilon, \varepsilon') = \frac{g_\mathrm{BCS}}{2\rf} \Theta(\debye-|\varepsilon-\Ef|)\Theta(\debye-|\varepsilon'-\Ef|),
\end{align}
where $g_\mathrm{BCS}$ is second order in the electron-phonon coupling $M$ and $\rf$ is the DOS at the Fermi energy. 
Throughout this article, we will use natural units $\hbar = \kb = 1$ and  $g_\mathrm{BCS}$ is chosen to be dimensionless.

However, even in second order in $M$, this expression appears to be a stark approximation. Avoiding it yields expressions, which depend on the difference $\varepsilon-\varepsilon'$ only. 
The simplest form {is rigorously derived via} a continuous unitary transformation based on an infinitesimal generator depending on the sign of 
$\varepsilon-\varepsilon' \pm \debye$ yielding
\begin{equation}
\label{eqn:g}
 g(\varepsilon, \varepsilon') = \frac{g}{2 \tilde{\rho}} \Theta(\debye - |\varepsilon - \varepsilon'|)
\end{equation}
in second order of the electron-phonon interaction without any further approximation \cite{krull2012}.
A sketch of the derivation is provided in the Supplemental Material \cite{supp}.
Here, we defined the dimensionless coupling strength $g$ and the average DOS around 
the Fermi edge $\tilde{\rho} \coloneqq 1/(2 \debye) \int_{\Ef - \debye}^{\Ef + \debye} \! \md \varepsilon \rho(\varepsilon)$
so that the formula can be employed also in case of divergences or other strongly fluctuating DOSes.
We stress that the dimensionless coupling $g$ can easily take values up to ${\cal O}(1)$ although it
results from a perturbative expansion in the electron-phonon coupling $M$.
The reason is that the expansion in $M$ requires
$M/\debye$ to be small, but does not prevent $M^2\tilde{\rho}/\debye$ to be of order unity or larger 
\cite{yuzbashyan2022}.

{Note, the extension of the BCS theory towards inclusion of retardation effects and Coulomb repulsion (Eliashberg theory)  still relies on the BCS assumption that the attractive part of the interaction involves quasiparticles in the vicinity of the Fermi level yet the interaction is repulsive outside of this region\cite{Carbotte1990}. In contrast, the energy-difference dependent interaction, which we will employ,  allows for attractive Cooper-pairing interaction due to phonons beyond the immediate vicinity of the Fermi level. This occurs because we used the originally derived form of the Fröhlich interaction that do not impose a cutoff energy at a certain distance from the Fermi level}.
For this choice, we show that the order parameter in the superconducting phase is finite even far away from the
Fermi energy, yet with decreasing modulus upon increasing energetic distance $|\epsilon-\Ef|$ 
\cite{lenz1996,mielke1997,althuser2025}.

An intriguing scenario with unexpected qualitative features arises if a significant accumulation, i.e., a peak,
of DOS is present at $\ep$ in the energetic proximity to the Fermi energy. The distance in energy from $\Ef$
does not need to be smaller than the Debye energy $\debye$, but $\ep$ can be several $\debye$ away for moderate values of
$g$. Then, we find a sudden increase of the superconducting order parameter $\Delta(\varepsilon\approx\ep)$. 
Correspondingly, it is also accompanied by a strongly enhanced mean-field critical temperature. The whole phenomenon appears as a function of temperature in two stages strongly resembling two consecutive phase transitions. 
It is heralded by the softening of an additional collective mode, which should be detectable spectroscopically. 
Additionally, we calculate the quasiparticle DOSes in the Supplemental Material \cite{supp}  to provide another readily measured quantity.

\begin{figure}
    \centering
    \includegraphics[width=0.98\columnwidth]{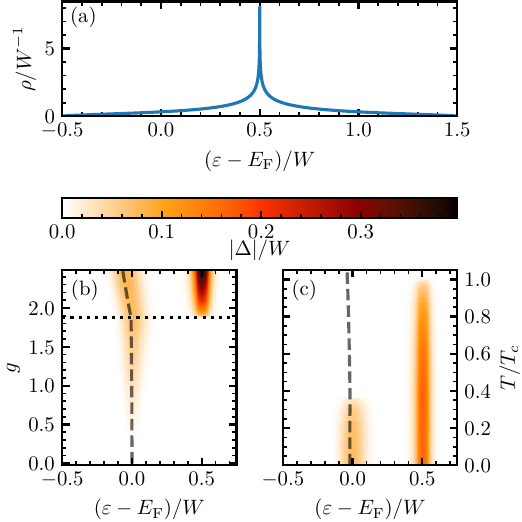}
    \caption{(a) Calculated DOS $\rho$ of a bcc lattice with nearest-neighbor hopping.
 (b) Order parameter $\Delta(\varepsilon)$ at $T=0$ vs.\ the interaction strength $g$ and the bare dispersion $\varepsilon$.
 The dotted line indicates the interaction strength $g_\mathrm{enh} \approx 1.86$ at which the sc order appears at the energy position of the logarithmic divergence.
 The dashed line marks the chemical potential $\mu(T)$.
 (c) Order parameter $\Delta(\varepsilon)$ vs.\ temperature $T$ at $g=2 > g_\mathrm{enh}$.
 For rising $T$, the order parameter first vanishes at $\Ef=-0.5W$, but remains finite at $\varepsilon=0$.
 At $T_c$, superconductivity vanishes.}
    \label{fig:gap_joined}
\end{figure}

%%%%%
% Model
%%%%%
Concretely, we study the Hamiltonian
\begin{align}
    \label{eqn:h}
 \hamiltonian &= \sum_{\vk \sigma} (\varepsilon_{\vk} - \mu) \ladder[\dagger]{\vk}{\sigma} \ladder{\vk}{\sigma} 
    \\ 
    &- \frac{1}{N} \sum_{\vk \vk' \sigma} g(\vk, \vk') \ladder[\dagger]{\vk}{\sigma} \ladder[\dagger]{-\vk}{-\sigma} 
 \ladder{-\vk'}{-\sigma} \ladder{\vk'}{\sigma}, \nonumber
\end{align}
where $\ladder[(\dagger)]{\vk}{\sigma}$ annihilate (create) an electron with momentum $\vk$ and spin $\sigma$, 
$\varepsilon_{\vk}$ is the single-particle dispersion, $\mu$ is the chemical potential, and $N$ the number of lattice sites.
We choose the hopping constant such that $\varepsilon\in [-W, W]$.
For simplicity, we assume a single optical phonon branch at the Debye frequency $\omega_{\vq} \equiv \debye$ 
and a constant electron-phonon coupling \cite{althuser2025}.
The chemical potential $\mu(T)$ is computed self-consistently such that the filling is kept constant.
We initialize the system with the Fermi energy $\Ef$ of the would-be normal state at $T=0$.
We fix $\debye=0.04W$ and $\Ef = -0.5W$ here, but the Supplemental Material \cite{supp} provides results for different parameters as well. 
We consider a generic Bravais lattice, here the body-centered cubic (bcc) lattice, with nearest-neighbor hopping.
This DOS  is plotted in \cref{fig:gap_joined}(a) \cite{joyce1971}.
A large fraction of its weight is located around the logarithmic divergence at $\vp=0=\ep$.
{We stress that a divergence is not required for the advocated phenomenon.
The only requirement is that a large weight is located beyond the Fermi level.
In the supplement, we examine such a DOS without a divergence and observe qualitatively the same features.}

We compute the static properties of the system, such as the heat capacity $C_V$ and the critical temperature $T_c$, by a standard self-consistent mean-field decoupling of the interaction term. To this end,
the DOS is discretized by an energy mesh with $10000$ equidistant points.
The expectation values $\langle\opSCd{\vk}\rangle$, and hence, the order parameter $\Delta(\varepsilon) = \int \md \varepsilon'  g(\varepsilon, \varepsilon') \rho(\varepsilon') \langle\opSCd{\vk}\rangle (\varepsilon')$, depend on the wave vectors only via the bare energy $\varepsilon$.
{To determine $T_c$, we solve the mean-field equations fully self-consistently starting at low temperatures and increasing $T$ until the order parameter vanishes.
Then, we fit the maximum of the order parameter to $\sqrt{T - T_c}$ for temperatures slightly below $T_c$ to obtain the proper value of $T_c$.}

%%%%%
% Results
%%%%%

\Cref{fig:gap_joined}(b) depicts the order parameter $\Delta(\vp)$ at  $T=0$.
The interaction strength $g$ is depicted along the $y$-axis, and $|\Delta|$ is shown by the color scale.
To guide the eye, the chemical potential $\mu$ is indicated by the dashed line.
Generally, $\mu(T) - \Ef$ is finite, but not particularly large.
Further discussions of $\mu$ as a function of $g$ and $T$ are provided in the Supplemental Material \cite{supp}.
For small to intermediate $g$, $\Delta$ is confined to the vicinity of the Fermi energy as expected in standard BCS theory.
But if $g$ is larger than some threshold value $g_\mathrm{enh}$, the order parameter becomes finite at $\ep$. 
This is induced by the enormous weight in the DOS around $\ep$.
{The large DOS causes the integral in the self-consistency equation
\begin{equation}
    \Delta(\ep) = g \int_{\ep - \debye}^{\ep + \debye} \md \vp' \rho(\vp') \frac{\Delta(\vp')}{2 E(\vp')}
\end{equation}
to become large even if $E(\vp')$ is large.
Note that this becomes relevant due to the consideration of an energy-transfer dependent interaction.}

The threshold $g_\mathrm{enh}$ depends on the specific model and its parameters.
Here, we find $g_\mathrm{enh} \approx 1.86$.

Fascinatingly, $\Delta(\ep)$ quickly surpasses $\Delta(\mu)$.
Yet, this peculiarity does not affect the energy gap: the minimum of the quasiparticle dispersion 
$E(\varepsilon) = \sqrt{(\varepsilon - \mu)^2 + |\Delta(\varepsilon)|^2}$ remains in the vicinity of the Fermi energy.
It may not be exactly at $\varepsilon=\mu$, but slightly shifted if $\Delta$ decreases quicker than $\varepsilon$ rises as discussed previously \cite{althuser2025}.
For clarity, we define $\trueGap \coloneqq \min_\varepsilon E(\varepsilon)$, i.e., the true energy gap, and 
$\maxGap \coloneqq\max_\varepsilon \Delta(\varepsilon) $, i.e., the maximum of the order parameter.
Furthermore, we refer to the order around $\mu$ as \emph{conventional SC order} and to the order at $\ep$ 
as \emph{enhanced SC order}.
For the present parameters, the conventional and enhanced $\Delta(\varepsilon)$ are not linked, i.e., between $\mu$ and $\ep$, there is an energy interval larger than $\debye$ in which $\Delta(\varepsilon)$ decreases exponentially so that it seems to vanish in the numerics.
Yet, we observed that their relative phase cannot be chosen arbitrarily, in particular if effects of the Coulomb interaction are included as well, see Supplemental Material \cite{supp}.

If the order parameter is tracked for $g < g_\mathrm{enh}$, the conventional BCS behavior is found except for some 
slight renormalization. For $g > g_\mathrm{enh}$, the enhanced order dominates.
Generally, the conventional order vanishes at a significantly lower temperature than the enhanced one, see
 \cref{fig:gap_joined}(c) for $g=2$. Thus, a strong effect in the critical temperature $T_c$ can be anticipated.
In this mean-field theory, the vanishing of the order parameter at the Fermi energy implies that the system becomes gapless,
though still in a superconducting phase. We do not, however, advocate the existence of such a phase 
in real systems because further interactions, such as the Coulomb repulsion, are known to connect the order parameter 
over all energies \cite{morel1962,mielke1997,sigrist2005,kostrzewa2018,simonato2023,althuser2025}.
{This view is corroborated by results, including a repulsive Hubbard-like  potential $U$, see the Supplemental Material \cite{supp}.
Of course, there is no argument that the Coulomb interaction would reduce to such a constant in the present model.
However, the inclusion of $U$ allows us to gauge how the enhancement reacts to additional interactions.
The consideration of the proper Coulomb interaction is an intriguing route for future research.}

\begin{figure}
    \centering
    \includegraphics[width=0.98\columnwidth]{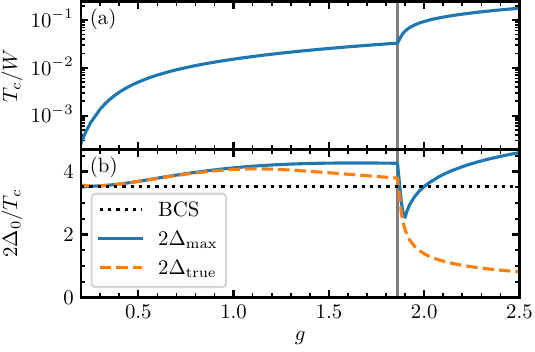}
    \caption{(a) Calculated critical temperature $T_c$ vs.\  the interaction strength $g$; note the logarithmic scale
 on the $y$ axis. From $g=g_\mathrm{enh} \approx 1.86$ (grey vertical line) onwards, the enhanced order significantly 
 increases $T_c$.
 (b) Ratio of the maximum value of the order parameter $\maxGap \coloneqq \max_\varepsilon |\Delta(\varepsilon)|$ (blue solid line) and the energy gap of the quasiparticle dispersion $\trueGap \coloneqq \min_\varepsilon E(\varepsilon)$ (orange dashed line) at $T=0$ relative to $T_c$ vs.\ $g$.
 The black dotted line marks the well-known ratio of standard BCS theory ($\approx 3.528$) \cite{czycholl2008}.}
    \label{fig:T_c}
\end{figure}

Indeed, the enhanced SC order has substantial effects on the critical temperature $T_c$.
In \cref{fig:T_c}(a),  $T_c$ is displayed against the interaction strength $g$.
For small $g$ values, there is only conventional order and $T_c$ rises slowly with $g$ 
essentially displaying standard BCS behavior.
However, upon reaching $g_\mathrm{enh}$, $T_c$ rapidly increases, mimicking a phase transition.
However, there is no additional symmetry breaking.
Such behavior is known in certain magnetic systems, where it is called a metamagnetic transition \cite{wohlfarth1962}.
Similarly, one might call the transition here a metasuperconducting transition.
In \cref{fig:T_c}(b), the ratio of $\maxGap$ (blue solid line) and $\trueGap$ (orange dashed line) at $T=0$ are depicted
relative to $T_c$. For weak coupling, the BCS ratio $2 \Delta_0 / T_c \approx 3.528$ \cite{czycholl2008} is recovered, 
marked by the black dotted line. At intermediate coupling strengths, the ratio is slightly larger than the BCS prediction.
In parallel, $\trueGap$ and $\maxGap$ no longer coincide, see Ref.\ \cite{althuser2025}.

At $g=g_\mathrm{enh}$, both ratios plummet.
While the drop is very fast, the numerics do not indicate it to be discontinuous.
As $\maxGap = \Delta(\ep)$ rises quickly, the ratio $\maxGap / T_c$ recovers quickly.
But the energy gap itself is not enhanced, so that the ratio  $\trueGap / T_c$ remains at a low value.

\begin{figure}
    \centering
    \includegraphics[width=0.98\columnwidth]{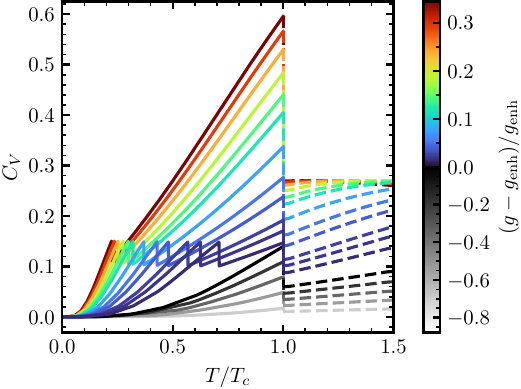}
    \caption{Calculated specific heat $C_V$ vs.\ temperature $T$.
 The colors represent the coupling strengths $g$ according to the color scale.
 Solid lines refer to the superconducting state, while dashed lines refer to the normal state.
 For $g < g_\mathrm{enh}$ (in grayscale), we observe renormalized BCS behavior.
 For $g > g_\mathrm{enh}$ (colored), unexpected qualitative behavior arises resembling two consecutive
 phase transitions.
 The threshold value is $g_\mathrm{enh} \approx 1.86$.}
    \label{fig:C_V}
\end{figure}

\Cref{fig:C_V} shows the specific heat $C_V = \partial \expec{\hamiltonian} / \partial T$.
Manifestly, the system exhibits superconductivity for $T < T_c$, indicated by the solid lines.
For $T > T_c$, the system is in the normal phase without order (dashed lines).
The color scale is twofold. The grayscale represents the data for $g<g_\mathrm{enh}$ displaying essentially
conventional BCS behavior: 
$C_V$ rises continuously upon increasing temperature, then jumps at $T_c$ as the system enters the normal phase.
The color scale represents the data for $g > g_\mathrm{enh}$, where the specific heat initially rises with rising temperature, then discontinuously drops, but rises again
and drops again when $T_c$ is reached.
The first drop concurs with the vanishing of $\Delta(\mu)$, i.e., the order parameter at the Fermi energy.
Considering larger $g$ or larger $\debye$ causes the conventional and enhanced order parameters to connect (not shown), 
i.e., there is no energy range between the two features in which the order parameter vanishes.
Then the first discontinuity in $C_V$ appears smeared out, resembling a $\lambda$-transition; see the Supplemental Material \cite{supp}.
The smearing-out increases as the energetic overlap between the two features increases. 
{Including the Hubbard-like repulsion interaction $U$ also implies that the discontinuity at lower temperatures is rounded.}

Finally, we also address spectroscopic signatures linked to the scenario of enhanced SC in order to extend the set of signatures of the advocated scenario.
To this end, we consider the operators
\begin{subequations}
\begin{align}
\label{eqn:response_ops}
 \mathfrak{A}_\mathrm{Higgs} &=   \frac{1}{\sqrt{N}} \sum_{\vk} \left( \opSCd{\vk} + \opSC{\vk} \right) \\
 \mathfrak{A}_\mathrm{Phase} &= \frac{\im}{\sqrt{N}} \sum_{\vk} \left( \opSCd{\vk} - \opSC{\vk} \right),
\end{align}
\end{subequations}
which are designed to excite Higgs modes or phase modes, respectively \cite{althuser2024}. We evaluate the spectral 
functions $\spectralMatSymbol_\alpha (\omega) = -(1/\pi) 
\Im \left[ \greenMatSymbol_{\mathfrak{A}_\alpha \mathfrak{A}_\alpha^\dagger} (\omega) \right]$ 
with $\alpha \in \{ \mathrm{Higgs}, \mathrm{Phase} \}$ at zero temperature via iterated equations of motion \cite{uhrig2009,kalthoff2017,bleicker2018,schwarz2020a,althuser2024}
truncated to bilinear operators. This corresponds to the solution of the Bethe-Salpeter equation with diagrams of bare propagators
of the mean-field Hamiltonian with exactly two fermionic propagator lines and all conceivable interaction lines of arbitrary number.
Experimentally, there is no direct coupling of light to the Higgs mode \cite{pekke15} so that its observation
is challenging. It can succeed in the nonlinear regime when the full Mexican hat potential is shaken \cite{schny11,krull14,schwa20}.

\begin{figure}
    \centering
    \includegraphics[width=0.98\columnwidth]{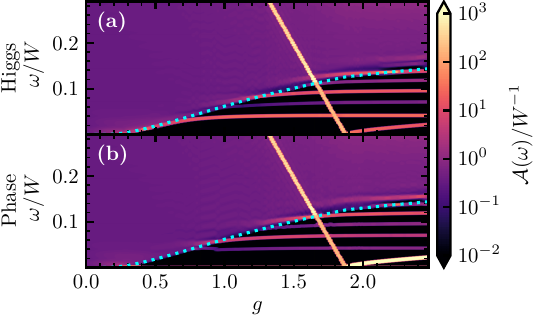}
    \caption{Calculated spectral functions of  (a) $\spectralMat{Higgs}$ and (b) $\spectralMat{Phase}$ represented by the color scale.
 The energy varies along the $y$-axis, and the interaction strength $g$ along the $x$-axis.
 The cyan dotted line marks the lower edge of the quasiparticle continuum $2 \trueGap$.
 The secondary modes found in Ref.~\cite{althuser2025} are clearly visible.
 Additionally, a sharp mode rapidly decreases in energy until softening at $g=g_\mathrm{enh} \approx 1.86$.
    $\delta$-peaks below the continuum are shown as Gaussian curves of the same weight and width $\sigma=0.0005W$.
 The primary phase mode appears at zero energy as a $\delta'$-peak since no long-range Coulomb
 interaction is included; it is plotted as the derivative of a Gaussian curve.
 The evaluation uses $16000$ equidistant points in $\varepsilon$-space.
 }
    \label{fig:resolvents}
\end{figure}

In \cref{fig:resolvents}, we plot the spectral functions (a) $\spectralMat{Higgs}$ and (b) $\spectralMat{Phase}$ 
 varying the interaction strength $g$ on the $x$-axis.
Collective excitations at finite energy within the gap are mathematically $\delta$-peaks.
We represent them by Gaussian curves with the same weight and the artificial broadening  $\sigma=0.0005W$.
The Goldstone boson associated with the phase fluctuations of the order parameter appears as a 
$\delta'$-peak and thus is plotted as the derivative of a Gaussian.
We point out that we have not included a long-range interaction, such that the energy of the Goldstone mode remains gapless, for explicit calculations of this effect see, e.g., Ref.\ \cite{althuser2025}.
First and foremost, we observe the primary Higgs mode slightly below the quasiparticle continuum for sufficiently strong coupling.
This process was first observed in Ref. \cite{barankov2007} using a phenomenological momentum-dependent interaction.
The same behavior was independently reported by Lorenzana and co-workers \cite{lorenzana2024,lorenzana2024a}.
At even larger $g$, secondary modes detach from the quasiparticle continuum \cite{althuser2025}.

In addition, there is another remarkably sharp mode appearing even inside the quasiparticle continuum.
Its energy rapidly decreases as $g$ increases; at $g=g_\mathrm{enh}$, it becomes soft.
This behavior typically signals quantum phase transitions, suggesting to interpret the appearance of $\Delta$ far away from the Fermi energy as another superconducting phase and the sharp mode (relative phase mode) as its precursor.
This interpretation is corroborated if one restricts the sum in the operator in \cref{eqn:response_ops} 
to a narrow energy interval $|\varepsilon(\vk)| < 0.01W$, i.e., the immediate vicinity of the peak in the DOS.
Then, the spectral functions do not show signatures of phase or Higgs modes, but only reflect the additional low-energy relative phase mode.
Hence, the additional mode is unambiguously linked to the enhanced superconductivity occurring at the peak in the DOS. 

{In this idealized setting, the enhanced order does not relate to the conventional order at all.
Therefore, the relative phase mode does not strongly couple to the bulk of quasiparticles, which continue to live around the Fermi level.
In contrast, the quasiparticle states related to the enhanced order have energies comparable to $\Ef$.
Thereby, the mode can stay sharp even inside the continuum.
This statement is corroborated in the Supplemental Material \cite{supp}, where the mode becomes much less sharp, if the enhanced and conventional orders overlap, either by shifting $\Ef$ or considering additional interactions.}

%%%%%
% Conclusion/Discussion
%%%%%
To conclude, we found enhanced and dominant SC order at an energy $\ep$ beyond the Fermi level for sufficiently strong coupling if the DOS exhibits a substantial accumulation of weight at this energy. 
This peak in the DOS does not need to be within the energy range set by the phonon frequency $\debye$, i.e., $|\Ef-\ep|> \debye$ is no showstopper.
Therefore, this enhancement is not captured by the conventional BCS approximation.
The enhanced SC order changes the system's thermodynamic properties substantially.
The critical temperature $T_c$ drastically increases if the coupling strength exceeds a threshold value $g_\mathrm{enh}$, depending on the DOS and the phonon frequency. 
Counterintuitively, this implies a lower value of the ratio $2\trueGap / T_c$.

In the specific heat, the enhanced superconductivity adds a jump-like feature which is reminiscent of a second phase transition, i.e., one transition where the enhanced order parameter at $\ep$ sets in and one where the conventional order at $\Ef$ sets in. 
Consistent with this view, we find an additional collective mode (relative phase mode), which softens where the enhanced superconductivity emerges. 
This mode is sharp even though it overlaps with the two-quasiparticle continuum.

Both signatures, the one in the specific heat and the one in the spectral response, will allow one to search experimentally for signatures of the theoretically identified enhanced superconductivity. 
In this quest, one must keep in mind that our calculations up to now do not account for additional interactions such as the long-range Coulomb interaction. 
{On the level of a local Hubbard-like repulsion, we find the same qualitative features in specific heat and in the spectroscopic response, see Supplemental Material \cite{supp}.}
But the features become smoother, i.e., jumps become steep increases, but they are no longer discontinuous.
The collective mode acquires a certain width, i.e., it is no longer as sharp as before. 
Its energy diminishes, but does not necessarily vanish completely.

Yet we emphasize that the phenomenon itself, enhanced superconductivity away from the Fermi level, does persist. 
This opens an intriguing research perspective for follow-up studies, both theoretically and experimentally.

%\begin{acknowledgments}
\textit{Acknowledgements --} We acknowledge constructive discussions with D. Hering, V. Sulaimann, S. Behrensmeier, and J. Stolze.
This research was partially funded by the MERCUR Kooperation in
project KO-2021-0027.
%\end{acknowledgments}

\nocite{*}

\bibliography{EnhancedSC.bib}

\end{document}